\documentclass[prl,twocolumn,showpacs,preprintnumbers,amsmath,amssymb,superscriptaddress]{revtex4}
\usepackage{epsfig}
\usepackage{subfigure}
\usepackage{graphicx}
\usepackage{dcolumn}
\usepackage{bm}
\usepackage{epsfig}

\def\e{\begin{equation}}
\def\f{\end{equation}}
\def\=#1{\overline{\overline{#1}}}
\def\_#1{{\bf #1}}

\def\.{\cdot}

\begin{document}

\preprint{APS/123-QED}

\title{A Numerical Study of Evanescent Fields
in Backward-Wave Slabs}

\author{M.K. K\"arkk\"ainen}
\affiliation{Radio Laboratory, Helsinki University of Technology,
P.O. Box 3000, FIN-02015 HUT, Finland} \email{mkk@cc.hut.fi}
\author{S.A. Tretyakov}
\affiliation{Radio Laboratory, Helsinki University of Technology,
P.O. Box 3000, FIN-02015 HUT, Finland}
\author{S.I. Maslovski}
\affiliation{Radio Laboratory, Helsinki University of Technology,
P.O. Box 3000, FIN-02015 HUT, Finland}
\author{P.A. Belov}
\affiliation{Radio Laboratory, Helsinki University of Technology,
P.O. Box 3000, FIN-02015 HUT, Finland} \affiliation{Physics
Department, St. Petersburg Institute of Fine Mechanics and Optics,
Sablinskaya 14, 197101, St. Petersburg, Russia}


\date{\today}

\begin{abstract}
Numerical study of evanescent fields in an isotropic backward-wave (BW)
(double negative or ``left-handed") slab is performed with the FDTD method.
This system is expected to be able to restore all spatial Fourier components of the
spectrum of a planar source, including evanescent fields, realizing a
``superlens".
The excitation of surface modes on the interfaces of the slab, which is the
key process responsible for the sub-wavelength focusing,  is
numerically confirmed and the time-domain field behavior is studied. In
particular, a numerical verification of the amplification of
evanescent modes by an isotropic BW slab with
$\epsilon_r=\mu_r=-1$ is provided.
\end{abstract}

\pacs{02.70.Bf, 
41.20.Jb, 
77.22.Ch, 
77.84.Lf 
}




\maketitle


The resolution wave limit of any optical device is well known:
it is impossible to resolve details smaller than half wavelength.
The physical reason of this limitation comes from the fact that evanescent
waves in the Fourier spectrum of an object exponentially
decay in the direction from the object plane. The decay factor reads
$\alpha=\sqrt{k_t^2-k_0^2}$, where $k_0$ is the wavenumber in free space,
and $k_t$ is the wavenumber of Fourier components in the object plane.
The faster the field varies in the object plane, the faster it
decays  in the direction normal to the object plane.
However, it was recently found that a very special kind of lens
made of a material with negative relative parameters $\epsilon_r=-1, \mu_r=-1$
(at a certain frequency) can ``amplify" the evanescent part of the
spectrum, thus opening a way to realize a
``superlens" with sub-wavelength resolution \cite{Pendry,Pendry2,Ramakrishna}.

The reason for this counterintuitive behavior is the fact that for a fixed frequency,
an interface between free space and
a backward-wave medium with  $\epsilon_r=-1, \mu_r=-1$
supports surface waves with arbitrary propagation constants
along the interface \cite{mine}.
Indeed, the eigenvalue equation for surface modes (surface polaritons)
on an interface between two isotropic media with  parameters
$\epsilon_{1,2}$ and $\mu_{1,2}$ reads
\begin{eqnarray}
{\sqrt{k_1^2-k_t^2}\over{\epsilon_1}}+{\sqrt{k_2^2-k_t^2}\over{\epsilon_2}}=0,\qquad
\mbox{TM modes} \label{TM} \\
{\mu_1 \over{\sqrt{k_1^2-k_t^2}}}+{\mu_2
\over{\sqrt{k_2^2-k_t^2}}}   =0,\qquad \mbox{TE modes} \label{TE}
\end{eqnarray} where indices $1,2$ correspond to the two media. If
$\epsilon_2=-\epsilon_1$ and $\mu_2=-\mu_1$, both \ref{TM} and
\ref{TE} are identically satisfied for all propagation constants
$k_t$ along the interface. This means  that any evanescent
incident plane wave is exactly in phase with one of the eigenmodes
of the surface wave spectrum. If the interface is infinite in
space, the amplitude of the excited surface wave becomes infinite.
The amplification of evanescent fields in backward-wave slabs
utilizes this resonant excitation of waveguide modes with large
propagation constants. The incident field excites an eigenmode of
the slab that is formed by two exponentially decaying field
components inside the slab. The spectrum of eigenmodes traveling
along a slab can be found from the expression for the reflection
or transmission coefficients: they have singularities at the
spectral points. For material parameters satisfying $\mu_r=-1$ and
$\epsilon_r= -1$ the transmission coefficient equals simply $\exp
(\alpha d)$, where $\alpha$ is the decay factor of an evanescent
mode from the source and $d$ is the slab thickness \cite{Pendry}.
Thus, there is only one eigenmode, and that solution  corresponds
to infinitely large $k_t$ and $\alpha$.  For larger  values of
$\alpha$ the excitation is closer to the resonance with this
eigenmode, and the field amplitude excited in the slab waveguide
is larger.

In paper \cite{Garcia}, however, it has been concluded that
if the width of the slab is limited,  the restoration of fields is physically
meaningless as it involves infinite energy.
To clarify the behavior of the evanescent fields in a BW slab or a finite
width, we
study the fields from a source that creates {\em only}
evanescent spectrum in the time domain with the finite-difference time-domain (FDTD)
method. Time
domain waveforms at suitably chosen observation points and
snapshot field distributions are calculated. Our simulations show that
the amplification indeed occurs,  and it is due to the surface modes
excited on the slab boundaries.


In any backward-wave medium the negative permittivity and permeability
must be dispersive, and here we adopt the
Lorentz model to account for the frequency dispersion.
The expressions for the permittivity and
permeability are of the form
\begin{eqnarray}
\epsilon(\omega) & = & \epsilon_0 \left( 1+
\frac{\omega_{pe}^2}{\omega_{0e}^2-\omega^2+j\Gamma_e \omega}
\right), \nonumber \\
\mu(\omega) & = & \mu_0 \left( 1+
\frac{\omega_{pm}^2}{\omega_{0m}^2-\omega^2+j\Gamma_m \omega}
\right). \label{eq:perms}
\end{eqnarray}
Here, $\omega_{0e}$ and $\omega_{0m}$ are the resonant frequencies
of the material and $\Gamma_e$ and $\Gamma_m$ are the damping
factors, which are assumed to be zero in numerical simulations. This
model corresponds to a realization of BW materials as mixtures of
conductive spirals or omega particles, as discussed in
\cite{Tretyakov}. In this artificial material both electric and
magnetic polarizations are due to currents induced on particles of
only one shape, which provides a possibility to realize the same
dispersion rule for both material parameters, as in
(\ref{eq:perms}). Note that the medium realized by Smith {\it et.\
al.} is built using different principles \cite{Smith}.
Numerical techniques appropriate to simulations of a material with the above
parameters with FDTD can be found in
\cite{mkkmotl2,Taflove,Young}. The discretization scheme used in
this paper is described in detail in \cite{mkkmotl2}.

Let the interface between free space and a BW slab lie along
the $x$-axis. The excitation plane (a line
in our 2D cut) is located at a distance $d_s$ from the BW slab of
thickness $d$.
In order to be able to study the behavior of the evanescent field,
we excite a BW slab by a source which produces no
traveling waves  in the direction orthogonal to the
source plane. We choose the incident electric field
that depends on the
$x$-coordinate along the interface and on the time $t$ in the
following manner:
\begin{equation}
E(x,t) = e^{-\left( \frac{x-x_0}{x_d} \right)^2} r(t) \cos{
(k_x x-\omega_0 t)}. \label{eq:propsrc}
\end{equation}
The ramp function $r(t)$ increases smoothly from $0$ to $1$ over
about $50$ periods of the cosine function. It is very important
that the amplitude grows slowly enough; a rapid increase in the
amplitude does not yield a constant amplitude on the second slab
interface or in the image plane, because of a wide
frequency spectrum of the source.  $x_0$ is taken to be the
$x$-coordinate in the middle of the slab and $x_d$ determines the
rate of decay of the incident electric field amplitude as measured
from $x_0$. $\omega_0$ is the center angular frequency of the
excitation. Due to the finite simulation space, we have given a
spatial profile to the incident field to obtain a decaying
amplitude near the boundaries of the simulation space. The shape
of the profile is the normal distribution. The
incident fields propagate to $+x$-direction along the interface
but decay exponentially in the $y$-direction normal to the slab
boundaries. From the dispersion relation in free space,
$k_x^2+k_y^2=\omega^2/c^2$ we see that by choosing $k_x
> \omega_0/c$ ($k_y=0$) in (\ref{eq:propsrc}) we obtain exponentially decaying
fields away from the source in the $y$-direction. In our
numerical simulations, we use $k_x=13.62$ m$^{-1}$ ($k_0=9.17$ m$^{-1}$).

The problem space is two-dimensional, with the field components
$H_x, H_y$, and $E_z$. The peak of the incident spectrum is at
$\omega_{0}=2.75 \cdot 10^9$ rad/s, and the parameters in
(\ref{eq:perms}) are the following: $\omega_{0e}=\omega_{0m}=0.55
\cdot 10^9$ rad/s, $\omega_{pe}^2=\omega_{pm}^2=1.482 \cdot
10^{19}$ (rad/s$)^2$, $\Gamma_e=\Gamma_m=0$. With these choices,
we obtain $\epsilon(\omega)=\mu(\omega)$ for all $\omega$ and
$\epsilon(\omega)/\epsilon_0=\mu(\omega)/\mu_0=-1$ at
$\omega=\omega_{0}$. Absorbing boundary conditions (ABC) are used
to terminate the computational domain at the outer boundaries of
the lattice. For simplicity, we have used Liao's third order ABC
\cite{Liao}, although more sophisticated ABC's are available. The
use of usual ABC's requires a small gap between the outer boundary
of the computational space and the BW material slab, since the
Liao's ABC is not applicable to dispersive media. The reflection
from the slab ends is negligible. We  excite a slab of thickness
$7 \Delta y=9$ cm, carrying out the simulations in a $500 \times
60$ lattice. The source plane is located at a distance $d_s=3.5
\Delta y= 5.25$ cm from the boundary of the slab (tangential
magnetic field components are defined on the interfaces).
Determining the precise spatial profile of the source, we take
$x_0=250 \Delta x$ and $x_d=x_0/5$ in (\ref{eq:propsrc}).

\begin{widetext}

\begin{figure}[htb]
\mbox{\subfigure[]{\epsfig{figure=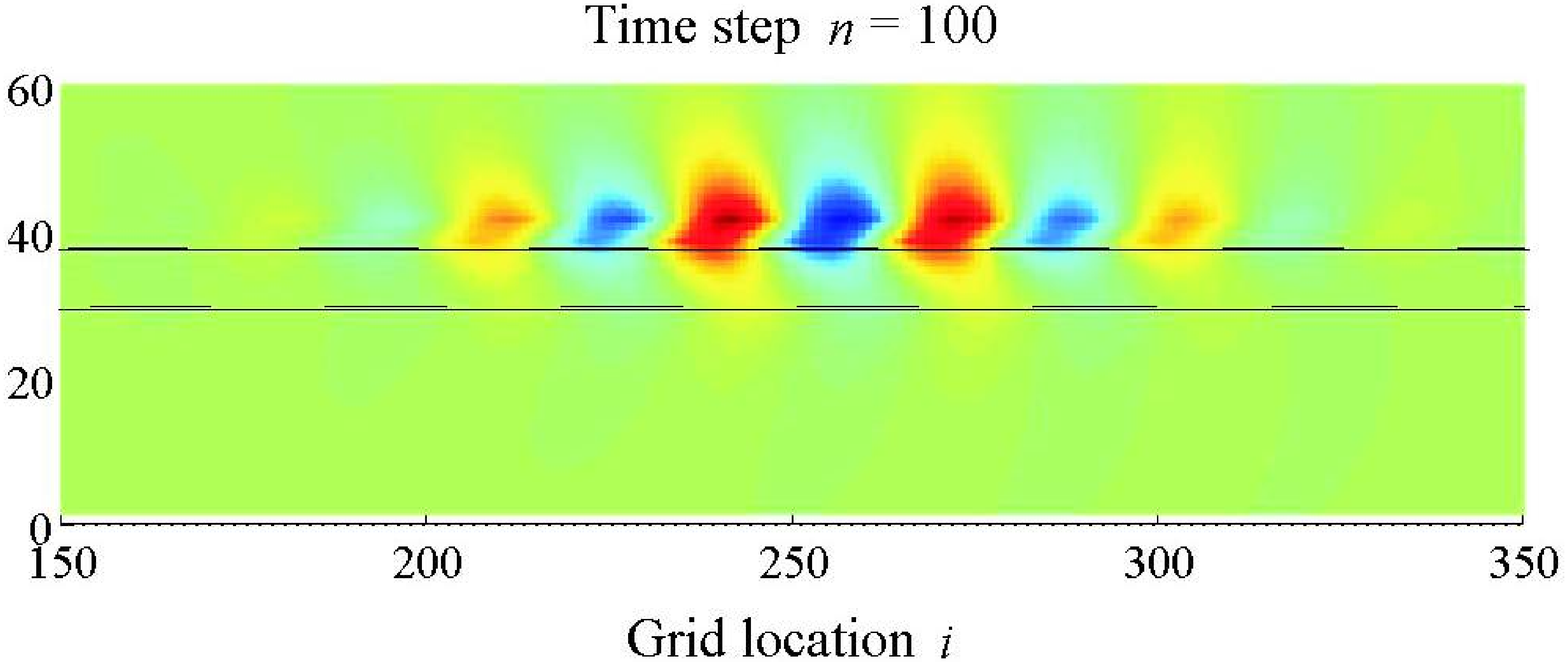,width=.32\textwidth,angle=0}}
\quad
\subfigure[]{\epsfig{figure=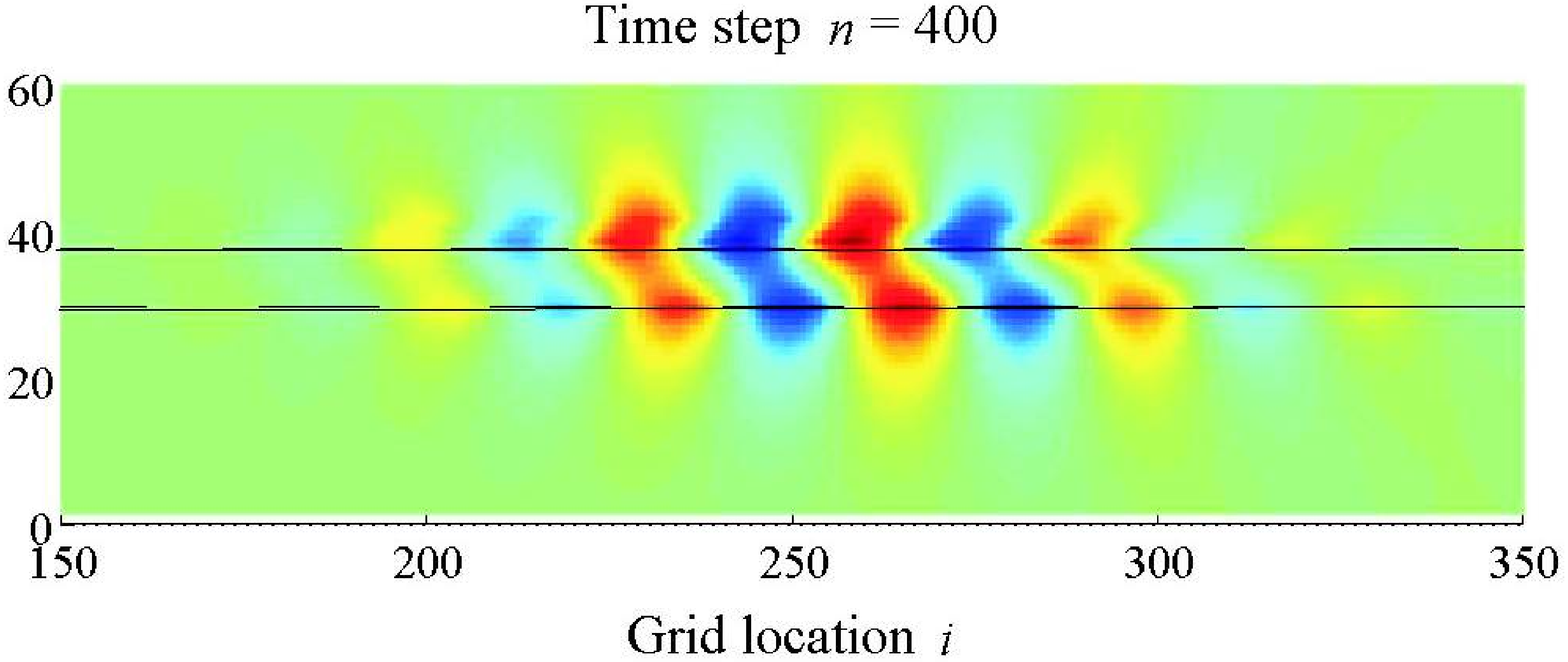,width=.32\textwidth,angle=0}}
\quad
\subfigure[]{\epsfig{figure=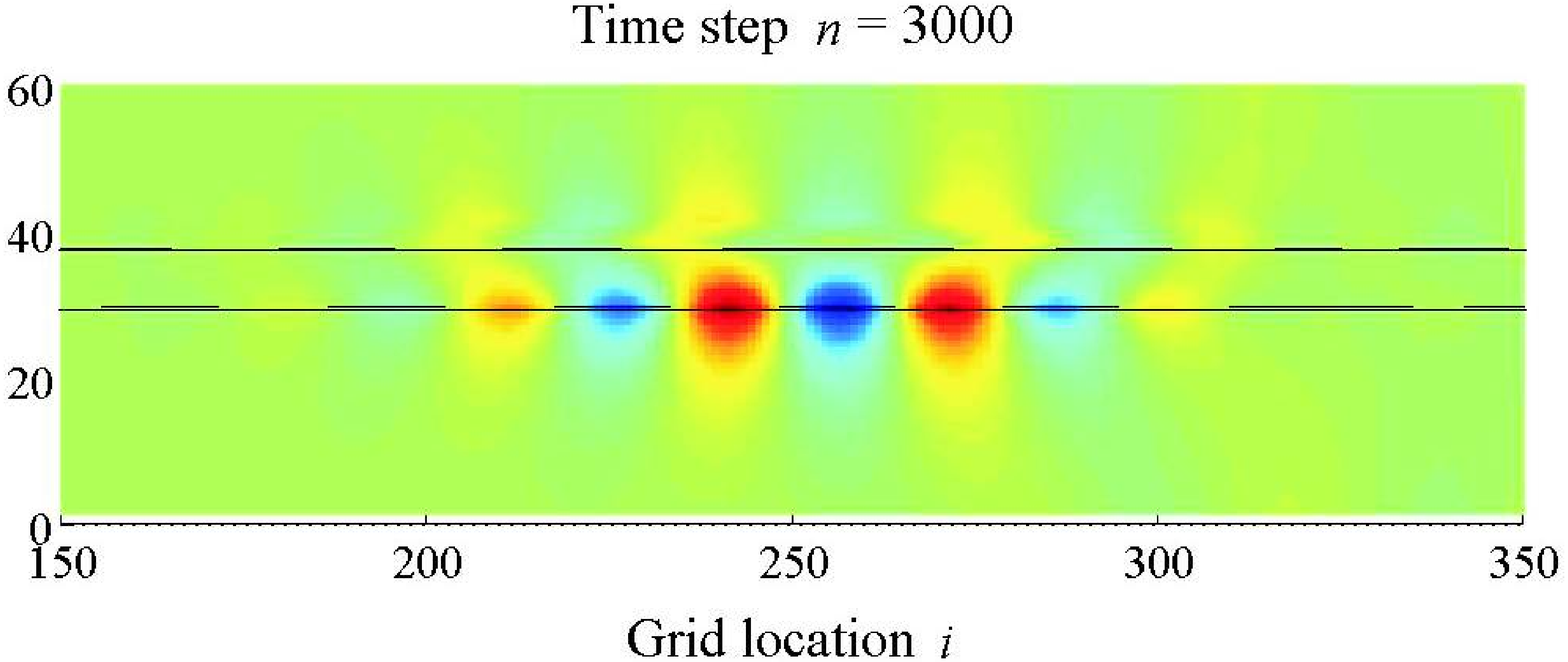,width=.32\textwidth,angle=0}}}
\caption{\small (a) The source fields are visible above the slab.
(b) The evanescent fields above the slab create surface waves on
the boundaries of the slab. (c) The evanescent mode has been
amplified by the slab: strong field amplitudes are visible on the
second interface. The fields created by the source are barely
visible above the slab, although they are just reaching the
maximum amplitude.} \label{kuva1}
\end{figure}

\end{widetext}

The problem formulation described above enables us to study how
the evanescent fields behave in a BW slab and how the surface
modes are excited. The numerical results are presented and
discussed next.

FDTD simulations have been run to see how the electric fields behave on
the slab boundaries and inside the slab. The electric fields
created by the source at an early stage of the simulation are
seen in Figure \ref{kuva1}~a). The snapshot electric field
distribution in Figure \ref{kuva1}~b) shows that the fields have
passed through the slab and the field amplitudes on the
slab boundaries are high. The electric field distribution in Figure
\ref{kuva1}~c) is recorded at a later moment of time. Figure
\ref{kuva1}~c) reveals that the electric fields are much stronger
on the second interface than on the first interface. Evidently,
the excited evanescent mode is amplified by the slab. Notice that
the source is just reaching its maximum amplitude at $n=3000$,
although the fields created by the source are not visible in
Figure \ref{kuva1}~c) due to the scaling.

\begin{figure}[htb]
\centering \epsfig{figure=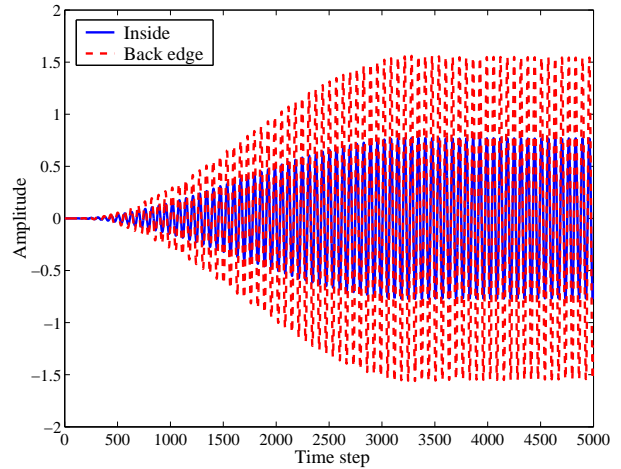, width=0.45\textwidth}
\caption{The electric field amplitude on the second boundary of
the slab is larger than that of the source field, and also larger
than the amplitude in the middle of the slab. This result verifies
that the evanescent fields are amplified by the slab.} \label{kuva2}
\end{figure}

\begin{figure}[htb]
\centering \epsfig{figure=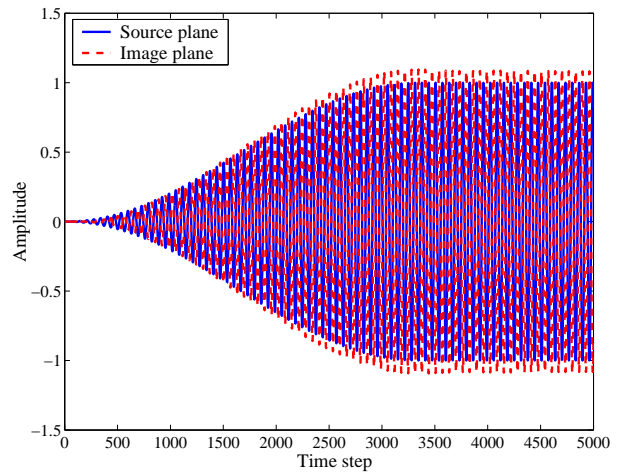, width=0.45\textwidth}
\caption{The evanescent electric fields in the image plane are close to the
fields in the source plane. This is also in agreement with the
theory.} \label{kuva3}
\end{figure}

Next we record the electric fields as a function of time at some
observation points. All the observation points are on the line
$x=250 \Delta x$: one in the middle of the slab, another on the
lower interface and the third in air in the image plane, which is located
at the distance $d-d_s$ from the second interface. The electric
fields at these points as functions of time are shown in Figures~\ref{kuva2}
and \ref{kuva3}. Figure~\ref{kuva2} shows that the
electric field amplitude is larger on the lower boundary of the
slab than inside the slab. The incident electric field amplitude
in the source plane is equal to unity. The source and the
observation point inside the slab are located at equal distances
from the upper slab boundary. Therefore, it is expected that the
electric field amplitude in the middle of the slab is equal to
$1$. We have observed a smaller amplitude in the middle of the slab. The
incident electric fields and the fields in the image plane are
compared in Figure~\ref{kuva3}. From the theory, is it expected
that the electric field amplitude in the image plane equals the
incident electric field amplitude. The fields in the source plane
and in the image plane are quite close to each other, see Figure~\ref{kuva3}.

Let us now compare the results in Figure~\ref{kuva2} with
analytical results. The evanescent mode decays away from the
source plane as a function of distance $y$ by a factor of
\begin{equation}
T_1 = e^{-\sqrt{k_x^2-k_0^2} y} \label{eq:decfac}
\end{equation}
until it hits the upper boundary of the slab. In the slab of
thickness $d$, the fields are amplified by a factor of
\begin{equation}
T_2 = e^{\sqrt{k_x^2-k_0^2} d}. \label{eq:ampfac}
\end{equation}
Combining these factors, we obtain the transmission coefficient
from the source plane to the second interface as
\begin{equation}
T = e^{\sqrt{k_x^2-k_0^2} (d-d_s)}, \label{eq:trans}
\end{equation}
where $d_s$ is the distance of the source from the upper boundary
of the slab. Substituting the parameters we obtain $T \approx
1.69$. In the numerical simulations, the
maximum ratio of the field amplitudes on the lower
interface and on the source plane equals $1.56$, being of the same order as
the analytical result. The fact that the calculated value is smaller should
have been expected, because in the numerical model  the incident field
amplitude decays from the slab center, while the estimation is for the
plane-wave excitation.

We have numerically demonstrated that the evanescent modes are
amplified in a lossless frequency dispersive backward-wave  slab.
The field amplitude at the other side of the slab is larger than the
source amplitude, and the ``amplification ratio" agrees well with the
analytical estimation. The  numerically
observed electric fields in the image plane
approximately reconstruct the source fields above the slab.
Thus, our numerical results support the conclusion of
paper \cite{Pendry} based on the theoretical analysis of
an infinite lossless backward-wave slab. The conclusion of \cite{Garcia}
about infinite field energy required for the restoration of
evanescent fields in finite slabs has no theoretical ground
(since the slab dimensions are finite and the field amplitudes are
finite everywhere if all the
wavenumbers in the spectrum are finite)
and is in conflict with the
results of numerical simulations.

\bibliography{unislab}

\begin{thebibliography}{11}
\expandafter\ifx\csname natexlab\endcsname\relax\def\natexlab#1{#1}\fi
\expandafter\ifx\csname bibnamefont\endcsname\relax
  \def\bibnamefont#1{#1}\fi
\expandafter\ifx\csname bibfnamefont\endcsname\relax
  \def\bibfnamefont#1{#1}\fi
\expandafter\ifx\csname citenamefont\endcsname\relax
  \def\citenamefont#1{#1}\fi
\expandafter\ifx\csname url\endcsname\relax
  \def\url#1{\texttt{#1}}\fi
\expandafter\ifx\csname urlprefix\endcsname\relax\def\urlprefix{URL }\fi
\providecommand{\bibinfo}[2]{#2}
\providecommand{\eprint}[2][]{\url{#2}}

\bibitem[{\citenamefont{Pendry}(2000)}]{Pendry}
\bibinfo{author}{\bibfnamefont{J.}~\bibnamefont{Pendry}},
  \bibinfo{journal}{Phys. Rev. Lett.} \textbf{\bibinfo{volume}{85}},
  \bibinfo{pages}{3966} (\bibinfo{year}{2000}).

\bibitem[{\citenamefont{Pendry and Ramakrishna}(2002)}]{Pendry2}
\bibinfo{author}{\bibfnamefont{J.}~\bibnamefont{Pendry}} \bibnamefont{and}
  \bibinfo{author}{\bibfnamefont{S.}~\bibnamefont{Ramakrishna}},
  \bibinfo{journal}{J. Phys. Condens. Matter} \textbf{\bibinfo{volume}{14}},
  \bibinfo{pages}{8463} (\bibinfo{year}{2002}).

\bibitem[{\citenamefont{Ramakrishna et~al.}(2002)\citenamefont{Ramakrishna,
  Pendry, Wiltshire, and Stewart}}]{Ramakrishna}
\bibinfo{author}{\bibfnamefont{S.}~\bibnamefont{Ramakrishna}},
  \bibinfo{author}{\bibfnamefont{J.}~\bibnamefont{Pendry}},
  \bibinfo{author}{\bibfnamefont{M.}~\bibnamefont{Wiltshire}},
  \bibnamefont{and} \bibinfo{author}{\bibfnamefont{W.}~\bibnamefont{Stewart}},
  \bibinfo{journal}{cond-mat/0207026}  (\bibinfo{year}{2002}).

\bibitem[{\citenamefont{Tretyakov
  et~al.}(2003{\natexlab{a}})\citenamefont{Tretyakov, Maslovski, Nefedov, and
  K{\"a}rkk{\"a}inen}}]{mine}
\bibinfo{author}{\bibfnamefont{S.}~\bibnamefont{Tretyakov}},
  \bibinfo{author}{\bibfnamefont{S.}~\bibnamefont{Maslovski}},
  \bibinfo{author}{\bibfnamefont{I.}~\bibnamefont{Nefedov}}, \bibnamefont{and}
  \bibinfo{author}{\bibfnamefont{M.}~\bibnamefont{K{\"a}rkk{\"a}inen}},
  \bibinfo{journal}{cond-mat/0212393}  (\bibinfo{year}{2003}{\natexlab{a}}).

\bibitem[{\citenamefont{Garcia and Nieto-Vesperinas}(2002)}]{Garcia}
\bibinfo{author}{\bibfnamefont{N.}~\bibnamefont{Garcia}} \bibnamefont{and}
  \bibinfo{author}{\bibfnamefont{M.}~\bibnamefont{Nieto-Vesperinas}},
  \bibinfo{journal}{Phys. Rev. Lett.} p. \bibinfo{pages}{207403}
  (\bibinfo{year}{2002}).

\bibitem[{\citenamefont{Tretyakov
  et~al.}(2003{\natexlab{b}})\citenamefont{Tretyakov, Nefedov, Simovski, and
  Maslovski}}]{Tretyakov}
\bibinfo{author}{\bibfnamefont{S.}~\bibnamefont{Tretyakov}},
  \bibinfo{author}{\bibfnamefont{I.}~\bibnamefont{Nefedov}},
  \bibinfo{author}{\bibfnamefont{C.}~\bibnamefont{Simovski}}, \bibnamefont{and}
  \bibinfo{author}{\bibfnamefont{S.}~\bibnamefont{Maslovski}},
  \emph{\bibinfo{title}{Advances in Electromagnetics of Complex Media and
  Metamaterials}} (\bibinfo{publisher}{Kluwer Academical Publishers},
  \bibinfo{year}{2003}{\natexlab{b}}), p.~\bibinfo{pages}{99}.

\bibitem[{\citenamefont{Smith et~al.}(2000)\citenamefont{Smith, Padilla, Vier,
  Nemat-Nasser, and Schultz}}]{Smith}
\bibinfo{author}{\bibfnamefont{D.}~\bibnamefont{Smith}},
  \bibinfo{author}{\bibfnamefont{W.}~\bibnamefont{Padilla}},
  \bibinfo{author}{\bibfnamefont{D.}~\bibnamefont{Vier}},
  \bibinfo{author}{\bibfnamefont{S.}~\bibnamefont{Nemat-Nasser}},
  \bibnamefont{and} \bibinfo{author}{\bibfnamefont{S.}~\bibnamefont{Schultz}},
  \bibinfo{journal}{Phys. Rev. Lett.} \textbf{\bibinfo{volume}{84}},
  \bibinfo{pages}{4184} (\bibinfo{year}{2000}).

\bibitem[{\citenamefont{Taflove and Hagness}(2000)}]{Taflove}
\bibinfo{author}{\bibfnamefont{A.}~\bibnamefont{Taflove}} \bibnamefont{and}
  \bibinfo{author}{\bibfnamefont{S.}~\bibnamefont{Hagness}},
  \emph{\bibinfo{title}{Computational Electrodynamics -- The finite-difference
  time-domain method}} (\bibinfo{publisher}{Artech House},
  \bibinfo{address}{Boston}, \bibinfo{year}{2000}).

\bibitem[{\citenamefont{Young and Nelson}(2001)}]{Young}
\bibinfo{author}{\bibfnamefont{J.}~\bibnamefont{Young}} \bibnamefont{and}
  \bibinfo{author}{\bibfnamefont{R.}~\bibnamefont{Nelson}},
  \bibinfo{journal}{IEEE Antennas Propag. Magazine}
  \textbf{\bibinfo{volume}{43}}, \bibinfo{pages}{72} (\bibinfo{year}{2001}).

\bibitem[{\citenamefont{K{\"a}rkk{\"a}inen and Maslovski}(2003)}]{mkkmotl2}
\bibinfo{author}{\bibfnamefont{M.}~\bibnamefont{K{\"a}rkk{\"a}inen}}
  \bibnamefont{and}
  \bibinfo{author}{\bibfnamefont{S.}~\bibnamefont{Maslovski}},
  \bibinfo{journal}{Microw. and Opt. Tech. Lett.} \textbf{\bibinfo{volume}{5}},
  \bibinfo{pages}{129} (\bibinfo{year}{2003}).

\bibitem[{\citenamefont{Liao et~al.}(1984)\citenamefont{Liao, Wong, Yang, and
  Yuan}}]{Liao}
\bibinfo{author}{\bibfnamefont{Z.}~\bibnamefont{Liao}},
  \bibinfo{author}{\bibfnamefont{H.}~\bibnamefont{Wong}},
  \bibinfo{author}{\bibfnamefont{B.-P.} \bibnamefont{Yang}}, \bibnamefont{and}
  \bibinfo{author}{\bibfnamefont{Y.-F.} \bibnamefont{Yuan}},
  \bibinfo{journal}{Sci. Sin. Ser. A} p. \bibinfo{pages}{1063}
  (\bibinfo{year}{1984}).

\end{thebibliography}

\end{document}